\documentclass[reprint,amsmath,amssymb,aps,prd,]{revtex4-1}

\usepackage{graphicx}
\usepackage{dcolumn}
\usepackage{bm}
\usepackage{hyperref}
\usepackage{comment}

\usepackage[hyperref,svgnames]{xcolor}

\newcommand{\gsim}{\lower.7ex\hbox{$\;\stackrel{\textstyle>}{\sim}\;$}}
\newcommand{\lsim}{\lower.7ex\hbox{$\;\stackrel{\textstyle<}{\sim}\;$}}

\def\mpl{M_{\rm pl}}

\newcommand{\eV}{{\, {\rm eV}}}

\def\beq{\begin{equation}}
\def\eeq{\end{equation}}
\def\bea{\begin{eqnarray}}
\def\eea{\end{eqnarray}}
\def\bitem{\begin{itemize}}
\def\eitem{\end{itemize}}
\newcommand{\bec}{\begin{center}}
\newcommand{\eec}{\end{center}}
\newcommand{\ba}{\begin{array}}
\newcommand{\ea}{\end{array}}

\def\bar#1{\overline{#1}}
\def\vev#1{\left\langle #1 \right\rangle}

\def\inv{^{\raise.15ex\hbox{${\scriptscriptstyle -}$}\kern-.05em 1}}
\def\lbar{{\lower.35ex\hbox{$\mathchar'26$}\mkern-10mu\lambda}} 

\def\om#1#2{\omega^{#1}{}_{#2}}

\def\to{\rightarrow}

\let\<=\langle
\let\>=\rangle

\let\+=\uparrow

\let\be=\beta

\let\om=\omega

\def\beqa{\begin{equation}\begin{aligned}}
\def\eeqa{\end{aligned}\end{equation}}

\begin{document}

\title{Axiverse Strings}

\author{John March-Russell}

\author{Hannah Tillim}

\affiliation{Rudolf Peierls Centre for Theoretical Physics, University of Oxford, Beecroft Building, Oxford OX1
3PU, United Kingdom}

\date{\today}

\begin{abstract}
If the QCD axion solves the strong CP problem then light axion-like-particles (ALPs) are expected to be ubiquitous in string theory - the string axiverse.  Such ALPs can be the QCD axion and constitute dark matter (DM) or radiation, quintessence, and lead to new forces.  String ALPs are also expected to give rise to a multiplicity of cosmologically important global axion strings.   We study the properties of these axiverse cosmic strings
including the vital effects of moduli stabilization, and find that the string cores provide `portals' to different decompactifications - to be precise, the cores explore the large K\"ahler or complex structure boundary of moduli space.   As usual for global strings the tension $T_1\sim \Lambda^2 \log(L\Lambda)$ with inter-string separation, $L$, while $\Lambda$ can be small $\ll \mpl$.  At long distances from the string there are potential new signatures involving variations in Standard Model (SM) parameters (Yukawa couplings, gauge couplings, masses) and equivalence principle
violations.
\end{abstract}

\maketitle

\section{\label{sec:intro}Introduction}

The strong CP problem is one of the most compelling motivations for physics beyond the SM.  
The most explored solution, the Peccei-Quinn (PQ) mechanism \cite{Peccei:1977hh}, implies the existence of a new light pseudo-scalar particle \cite{Weinberg:1977ma,Wilczek:1977pj}, the axion, $a(x)$.
In QFT models of axions \cite{Kim:1979if,Shifman:1979if,Zhitnitsky:1980tq,Dine:1981rt}, $a(x)$ is essentially the phase of field(s) $\Phi$ whose vacuum expectation value(s) (VEV)
spontaneously breaks the U(1) PQ symmetry, with $a(x)$ the associated
Nambu-Goldstone boson.  For a single PQ-breaking field we may write $\Phi(x) =
|\Phi(x)| \exp(i a(x)/f)$, where, in vacuum, $\vev{|\Phi(x)|} = f/\sqrt{2}$.  The scale $f$ defines the axion periodicity while $1/f$ parametrically sets the strength of axion interactions.  
Classically the axion action has a continuous shift symmetry $a(x)\to a(x)+ {\rm const}$, 
but this is explicitly broken by non-perturbative QCD dynamics to a discrete symmetry
$a\to a+2\pi n f$, $n\in Z$,  and a potential is generated such that
the axion VEV cancels the CP-violating ${\bar\theta}$-term.  

To solve the strong CP problem, however, the QCD contribution
must dominate all other sources of explicit PQ
breaking by $\gtrsim 10^{10}$\cite{Kamionkowski:1992mf,Holman:1992us,Barr:1992qq}.  This \emph{axion quality problem} is non-trivial in quantum gravity theories where there is strong evidence that global symmetries are necessarily explicitly broken \cite{Giddings:1987cg,Banks:1988yz,Coleman:1989zu,Banks:2010zn,Harlow:2018tng,Daus:2020vtf}.  
Favouring the PQ solution is the fact that (possibly heavy) axion-like-particles (ALPs) are ubiquitous in our best understood quantum gravity theory, string theory \cite{Witten:1984dg,Svrcek:2006yi,Arvanitaki:2009fg}. 
Importantly, solving the axion quality problem in string theory (so that the QCD axion is light) implies
there exist {many light ALPs} \cite{Arvanitaki:2009fg}: a string axiverse.  The origin and UV physics of such axions is quite different to traditional QFT axions. 

In addition to particle excitations of $a(x)$, 
one may also consider topologically non-trivial axion cosmic string solutions where $a(x)$ winds as some simple closed curve in physical space is traversed \cite{Vilenkin:1982ks,Vilenkin:2000jqa}.  Then, for QFT axions, $|\Phi(x)|$ necessarily has a zero at some location, and this core region explores and is sensitive to UV physics.  The tension, $T_1\sim f^2 \log(Lf)$, of such global strings IR diverges with the system size, $L$, and moreover they become the boundaries of axion domain walls once $a(x)$ acquires a mass, and are thus formally ``confined". Cosmologically, however, the inter-string separation provides an IR cutoff, and a network of such strings and domain walls is important for both relic axion DM production \cite{Davis:1986xc,Harari:1987ht,Battye:1993jv,Hiramatsu:2012gg,Gorghetto:2018myk} and the generation of stochastic gravitational 
wave backgrounds \cite{Hindmarsh:1994re,Vilenkin:2000jqa,Saikawa:2017hiv,Gorghetto:2021fsn,Chang:2021afa}.  They can also lead to a variety of other striking, potentially observable, phenomena \cite{Brandenberger:1996zp,Agrawal:2019lkr,Fukuda:2020kym,Agrawal:2020euj}.

While many aspects of string axions have been explored, to the best of our knowledge none have touched on the underlying physics of possible axiverse cosmic string solutions and the crucial role moduli stabilisation plays.  These issues are the subject of this paper.

\section{\label{sec:Axiverse}String Axiverse Recap}

ALPs arise from the $k$-form fields present in the underlying string theory, such as the 2-form, $B_2$, of heterotic
and type II,  or the type II RR $k$-forms $C_{k}$ (see Sec.\ref{sec:general_strings}).   
Vitally, a \emph{single} $k>1$ form leads to a \emph{multiplicity} of classically massless ALP candidates, $a^i$, determined by the topology of the 6d Calabi-Yau (CY) compactification, $Z$.  
For example,  if $\om_i$ is a basis for the $h^{1,1}$ complex (1,1) harmonic forms
(dual to closed 2-cycles) of $Z$ then $B_2$ gives $h^{1,1}$ 
potential 4d ALPs via $B_2 = b^i(x) \omega_i$.  Related arguments apply to other $k$-form fields and $k$-cycles.

As realistic compactifications are topologically rich, with $10^{\rm few}$ cycles, there are a similar number of potentially light 4d ALPs.   The presence of fluxes, orientifolds, and branes lifts a subset at tree level, but those that survive remain massless in perturbation theory.

The most attractive compactifications preserve ${\cal N}=1$ supersymmetry (SUSY) before moduli stabilisation, so the $a^i$ are accompanied by `saxions' $t^i$ - the scalar compactification moduli.  Before including the stabilisation potential the leading 4d effective action of the moduli fields $m^i = a^i(x) + i t^i(x)$ is
($m^{\bar i} = (m^{i})^*$, $\partial_{i} = \partial/\partial m^{i}$)
\beq
{\cal L} = G_{i{\bar j}}(m,{\bar m}) g^{\mu\nu} \partial_\mu m^{i}  \partial_\nu  m^{\bar j} 
\label{eq:action0}
\eeq
where $G_{i{\bar j}} = \partial_{i} \partial_{{\bar j}} K(m,{\bar m})$ with $K$ the K\"ahler potential.  The behaviour of $G_{i{\bar j}}$ at moduli space boundaries will be of special interest for the construction of axiverse strings.

A moduli potential vanishing at tree level is protected from loop
corrections but generated by non-perturbative effects, and, as seen in
realistic stabilisation mechanisms, at least some of these involve SUSY-breaking dynamics.   
The $t^i$ then receive masses, while the $a^i$, being pNGBs,
can be separately protected
and often receive \emph{hierarchically smaller} non-perturbative masses.  In fact, this hierarchy in
$t^j$ and $a^j$ masses \emph{must} be the case if axiverse ALPs are light enough ($\lesssim 10^{-3}\eV$) to be relevant for the strong CP-problem or light `field' DM.

\section{\label{sec:strings} Axiverse Strings: Basics}

\begin{figure}
	\includegraphics[width=0.8\linewidth]{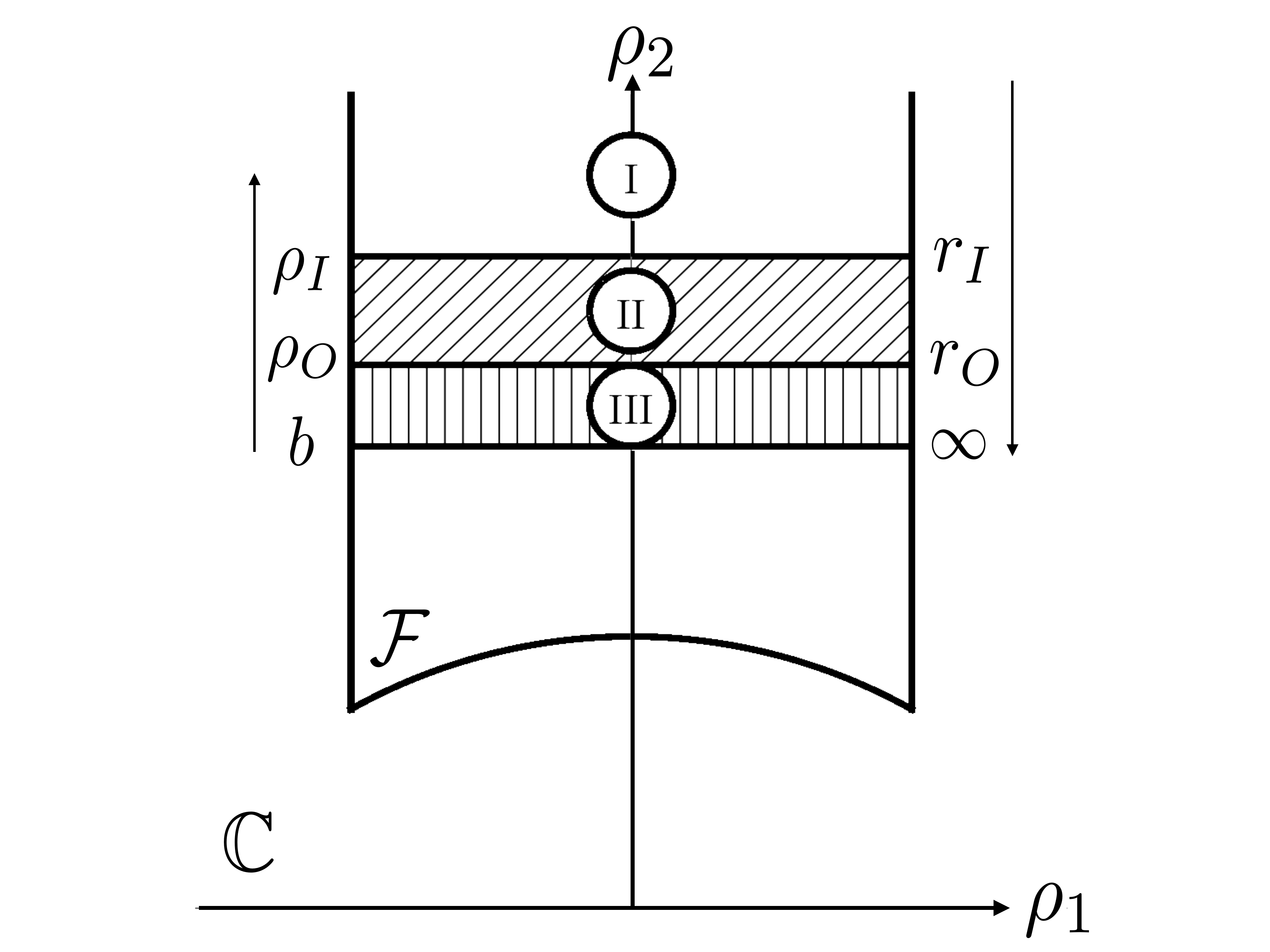}
	\caption{Domain ${\cal F}$ of $T^2$ K\"ahler moduli space with the regions 
		explored by I) inner core ($r<r_I$), II) outer core ($r_I \leq r \leq r_O$), III) far field ($r>r_O$), indicated.  
		} 
\label{fig:Kahler1}
\end{figure}

Consider the simple case of compactification on a 6-manifold of form $K_4\times T^2$ with ALP arising from the 2-form $B_{2}$ with legs in the $T^2$ directions, $y_m$, $m=1,2$
(we do not need to specify $K_4$).  If $h_{mn}$ is the $T^2$ internal metric, then
$\rho(x)\equiv B_{12} + i \sqrt{\det h}\equiv \rho_1+ i \rho_2$  is the dimensionless
complexified K\"ahler `size' modulus, and $\rho_1$ is our ALP of period $1$.
The domain, ${\cal F}$, of inequivalent $\rho$ fields is the upper-half ${\mathbb C}$-plane modulo $SL(2,\mathbb{Z})$ - see Fig.\ref{fig:Kahler1}.   
The reduction of the 10d string action gives $K(\rho,{\bar \rho})\propto\log(\rho_2)$, so $G_{\rho{\bar \rho}} \propto 1/4\rho_2^2$, and the relevant Einstein-frame 4d effective action of the light fields is
\beq
S = M^2\int d^4 x \sqrt{-\det g} \left\{ g_{\mu\nu} \frac{\partial^\mu \rho  \partial^\nu {\bar\rho}}{4\rho_2^2} -  V(\rho_2)   \right\}~.
\label{eq:action1}
\eeq 
Here the 4d dilaton $\varphi$ does not appear as we assumed it to be stiffly stabilised.  (In Sec.\ref{sec:GravityWarping}. we return to this and discuss the physics that sets the scale $M$.)
In Sec.\ref{sec:pheno} the effect of an additional non-perturbative potential for the axion is discussed assuming a well-separated hierarchy of scales $V(\rho_2) \gg  {\tilde V}(\rho_1)$, so $\rho_2$ is much heavier than $\rho_1$.  

A crucial feature of eq.(\ref{eq:action1}) is that the axion decay constant is
set by the partner VEV: $f_{\rm eff} = M/(\sqrt{2} \vev{\rho_2})$.
Thus, by analogy with usual QFT axion strings where $f_{\rm eff}(x) =\sqrt{2} \vev{|\Phi(x)|}$ has a zero at the string core, we expect \emph{axiverse cosmic strings to have cores where} $\rho_2\rightarrow \infty$, namely \emph{at a suitable boundary of the moduli space}.  This expectation is correct, and generalises to axiverse strings arising from compactifications other than a simple $T^2$.

\begin{figure}
	\includegraphics[width=0.9\linewidth]{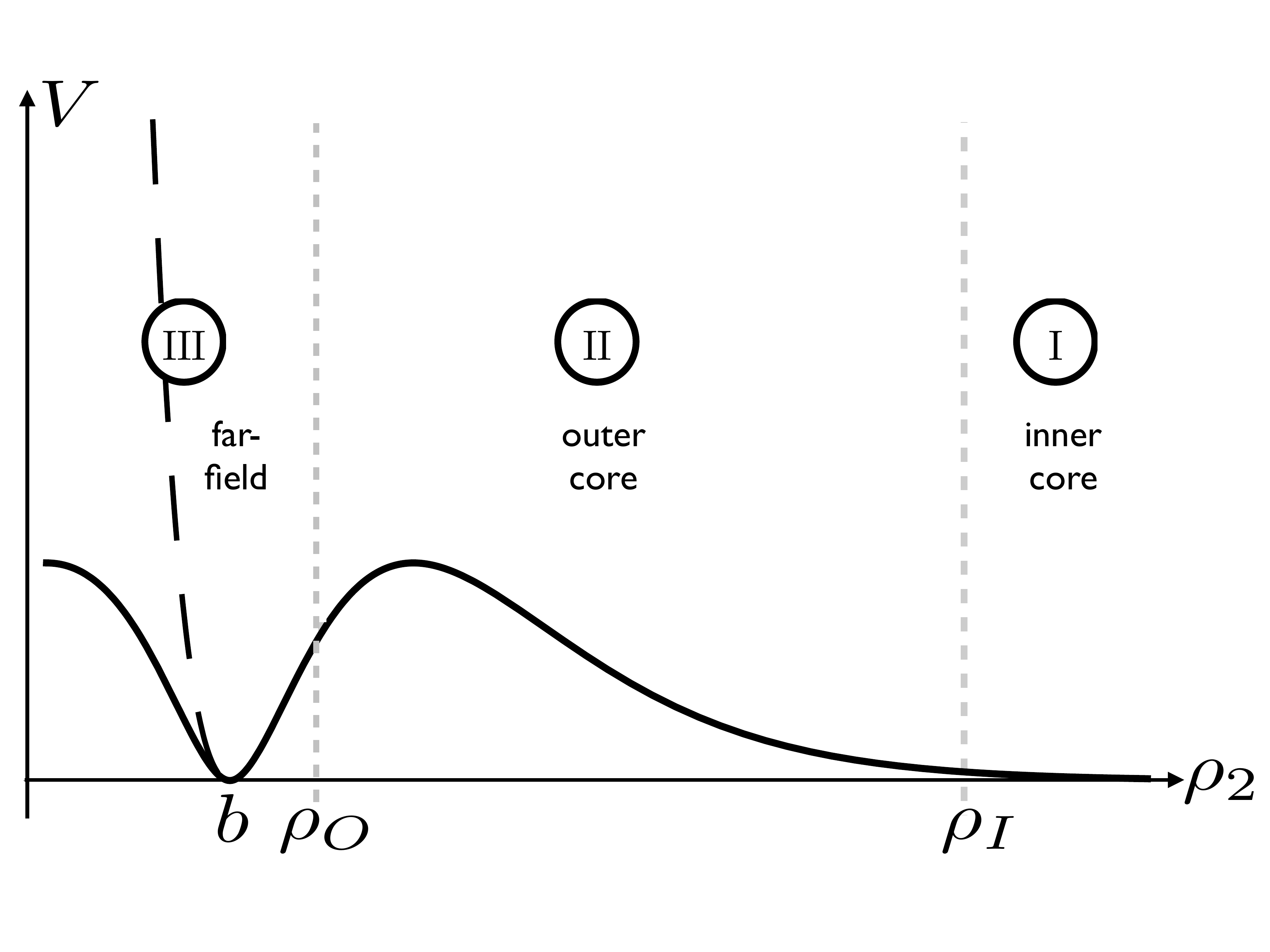}
	\caption{General shape of $V(\rho_2)$, with three regions delineated.  
		} 
\label{fig:Kahler2}
\end{figure}

Concerning $V(\rho_2)$, it must go to zero in the decompactification limit \cite{Giddings:2004vr,Ooguri:2006in,Blumenhagen:2018nts,Ooguri:2018wrx,Hebecker:2018vxz}, $\rho_2 \rightarrow \infty$, and, for stabilisation, have a local minimum at some $\rho_2=b>0$, with $V(b)\simeq 0$.  Additionally we require $b\gg1$ so that the size of the stabilised $T^2$ is large and 
the leading-order action, eq.(\ref{eq:action0}), is appropriate.  This is also necessary to sufficiently
suppress the non-perturbative effects that would otherwise lift all
ALP masses and eliminate the axiverse \cite{Arvanitaki:2009fg}.
We thus take $V(\rho_2)$ to be of the form shown in Fig.\ref{fig:Kahler2}.   

A toy-model potential with these properties is
\beq
V(\rho_2) = \mu^2 e^{-a \rho_2} (\rho_2 - b)^2 ~,
\label{Vofrho}
\eeq
where $\mu^2 e^{-a b} \ll M^2$ sets the scale.   
The canonical modulus mass is $m_c = 2b\mu e^{-ab/2}$ at $\vev{\rho_2}=b$.
The solution is indifferent to the shape of $V$ for $\rho_2<b$ as long as $b$ is a minimum, while the exponential decline of $V(\rho_2\rightarrow \infty)$ in eq.(\ref{Vofrho}) is for convenience - all results are qualitatively unaffected as long as $V(\rho_2\rightarrow \infty) \sim 1/\rho_2^n$ for $n>0$.

To proceed initially freeze the metric $g_{\mu\nu}(x)=\eta_{\mu\nu}$.
For an infinite string along the $x_3$-axis the
solution depends only on the transverse coordinates
$z\equiv x_1+i x_2\equiv r \exp(i \theta)$.   As analysed in Ref.\cite{Greene:1989ya} in the absence of $V$, the equation following from eq.(\ref{eq:action1}) is (here $\partial = \partial/\partial_z$ etc)
\beq
\partial {\bar \partial} \rho - \frac{2 \partial \rho {\bar \partial} \rho}{ \rho - {\bar \rho}} = 0~.
\label{eq:holoeqn}
\eeq
which is naively solved by any (anti)meromorphic $\rho$. 
However,  these solutions are not directly physically relevant as they are greatly 
altered by the potential.   Thus consider the {ansatz}
$\rho_1 = \theta/2\pi, ~\rho_2 = \rho_2(r)$
for the elementary winding number $w=1$ string.
This leads to
\beq
\frac{d^2 \rho_2}{d r^2}  + \frac{1}{r} \frac{d\rho_2}{d r} - \frac{1}{\rho_2} \left(\frac{d \rho_2}{d r}\right)^2 + \frac{1}{4 \pi^2 r^2 \rho_2} =  2 \rho_2^2 \frac{d V}{d \rho_2}~,
\label{eq:EoM}
\eeq
with $\rho_2(r\rightarrow 0)\rightarrow \infty$, and $\rho_2(r\rightarrow\infty) \rightarrow b$.
The solution is then found by stitching three
regions - see Figs..\ref{fig:Kahler1},\ref{fig:Kahler2}:\\
$~\quad$ Region I) Inner core: $0< r < r_I$, $\infty > \rho_2 > \rho_I$\\
$~\quad$ Region II) Outer core: $r_I\leq r \leq  r_O$, $\rho_I\geq \rho_2 \geq \rho_O$\\
$~\quad$ Region III) Far field: $r_O< r < \infty$,  $\rho_O > \rho_2 > b$\\ 
For $V(\rho_2)$ of eq.(\ref{Vofrho}) we take matching values $\rho_O \equiv b+1/a$ 
and $\rho_I \equiv b +8/a \gg 1$.   The features of the string solution do not depend on these specific choices.


In Region III $V$ is important, 
and the solution asymptotes to $\vev{\rho_2} = b$.  In the inner-core, where $\rho_2 \rightarrow \infty$, $V$
can be neglected.  In between there is an `outer core' transition region, which can be either `thick' or `thin'. 
In more detail, the $w=1$ solution in the $r>r_O$ region is
\beq
\rho_2 
= b + \frac{b}{(2\pi b\, m_c r)^2}  +\ldots ~,
\label{eq:farfieldsol}
\eeq 
with $1/r^2$ behaviour characteristic of global cosmic strings.  
Given $\rho_O\!=\! b+1/a$ the outer core radius is then
$r_O  \simeq (a/b)^{1/2}/2\pi m_c$.
In the inner core the RHS of eq.(\ref{eq:EoM}) may be neglected, giving an
approximate $w=1$ solution valid for $\rho>\rho_I= b +8/a \gg 1$,
\beq
\rho(z,{\bar z}) \simeq -\frac{i}{2\pi }\log(cz),~~c=\frac{\exp(-2\pi \rho_I)}{r_I}~.
\label{eq:innercoresol}
\eeq
(As the $\rho$ domain is ${\cal F}$, rather than a strip of $\mathbb{C}$ as naively given by eq.(\ref{eq:innercoresol}), in principle a more precise inner core solution is $j(\rho) = 1/cz$ where Klein's modular $j$-function maps the fundamental domain, ${\cal F}$, to $\mathbb{C}$ \cite{Greene:1989ya}.   However for \emph{physically realistic} strings with moduli stabilised the solution is only needed for $c|z|\ll 1$.   In terms of $q\equiv \exp(2\pi i \rho)$, $j(\rho) \simeq q^{-1} +744 + {\cal O}(q)$ for $|q| \ll 1$.  So in this limit
$\rho(z) \simeq -i\left(\log cz + 744\, cz  +\ldots \right)/2\pi$
agreeing with eq.(\ref{eq:innercoresol}) up to an error of
$<1\%$ for $\rho_2 > b > 2$ - though intriguingly slightly deviating from cylindrical symmetry as $\rho_1\simeq (\theta + 744\, cr \sin\theta + \ldots)/2\pi$.)  
Finally the numerically derived, non-universal, outer-core solution
stitches the above Region I and III solutions together.
 
The action, eq.(\ref{eq:action1}), implies
the $w=1$ tension $T_1$ is
\beq
2 \pi M^2 \!\!\! \int \!\! drr\! \left[ \frac{(\partial_r \rho_2)^2}{\rho_2^2}  \!+ \! \frac{1}{4 \pi^2 r^2 \rho_2^2}  \! +  \!\mu^2 e^{-a \rho_2} (\rho_2  \!- \! b)^2\right]
\label{eq:tension}
\eeq
Splitting this into contributions $T_{I,II,III}$ from the three regions, analysis shows that 
the far-field part $T_{III}$ is log-sensitive to the IR cutoff, $L$, while the leading inner-core
contribution gives $\Delta T_I = 2 M^2/\rho_I$,
independent of $r_I$ and \emph{finite} despite the decompactification.  Minimising $T_1$
with respect to $r_I$ for fixed $L, r_O$ shows that the thin wall case requires $a\gg 14\pi$.  
In the end, for both thin and thick wall cases, the tension at distance $r>r_O$ is
\beq
T_1(r) \simeq \frac{M^2}{2\pi b^2} \log \left( \frac{r}{r_O}\right) + {\cal O}\left(\frac{2M^2}{b}\right)~,
\label{eq:T1tension}
\eeq
similar to a normal axion global string.

Finally note that the action eq.(\ref{eq:action1}) and domain of the $T^2$ complex structure (CS) modulus
$\tau$ are identical to that for $\rho$, so these results apply to $T^2$ CS strings too.

\section{\label{sec:GravityWarping}Gravity and The Core}

For $r>  r_O$ the energy density of an axiverse string goes
as a conventional global string, with naively IR-divergent tension, eq.(\ref{eq:T1tension}).  For an isolated infinitely long static string this leads to a metric singularity at very large distances \cite{Cohen:1988sg}, but for the realistic situation of a string network
the inter-string separation gives a finite effective $L$, and the metric is smooth.   At intermediate distances, $r_O < r < L$, and in the observationally required limit $k(r) \equiv 4 G_N T_1(r)\ll 1$, the 4d metric is well approximated by
$ds^2\!\simeq\! (1\!-\! k(r)\!+\!k_0) (dt^2\! - dx^2_3) - dr^2\! - (1\!-\! 2 k(r)) r^2 d\theta^2$
implying both an $r$-dependent deficit angle $\Delta\theta \simeq 2\pi k(r)$ and
the well-known gravitational repulsion\cite{Harari:1988wa}.  The $V(\rho_2)$-dependent constant $k_0$ is
determined by numerically matching to the core solution.   

The mass scale, $M$, of the moduli
field action eq.(\ref{eq:action1}), equivalently the
axion decay constant, may be greatly reduced compared to $\mpl$ if warped compactifications 
\cite{Svrcek:2006yi,Flacke:2006ad,Randall:1999ee} are considered with the cycle on which the ALP is defined localised in the IR region with warp factor $e^{-w_0}\ll 1$.   $M$ may also be lowered in, eg, SM-on-a-brane large volume compactifications with large string length $\ell_s\equiv 2\pi\sqrt{\alpha'}$ \cite{Svrcek:2006yi}.  
Both warping and brane-world constructions allow $4 G_N T_1(r) \ll 1$ consistent with the gravitational constraints on cosmic strings (see e.g. \cite{LIGOScientific:2021nrg}).  

Turning to the inner core, the ansatz $ds^2_4 = dt^2 - dx^2_3 - p(z,{\bar z}) dz d{\bar z}$ does not alter eq.(\ref{eq:holoeqn}), so eq.(\ref{eq:innercoresol}) is still a solution in this region \cite{Greene:1989ya}. Analysis of the 4d Einstein equations shows that $p(z,{\bar z})\sim -\log(r)$ as $r\rightarrow 0$.  Importantly, since the decompactification is localised in 4d, $\mpl$ remains \emph{finite} and set by the asymptotic $\vev{\rho_2}=b$.  
Strictly speaking, though, the 4d effective theory breaks down for \emph{local} experiments in the inner core as $r\rightarrow 0$.  

A better description follows from a 6d theory with $\rho_1,\rho_2$ replaced by $B_2, h_{mn}$.  One then sees that the $B_2$ and metric configuration is due to an effective string in 6d oriented along the $x^3$ axis that magnetically sources $B_2$  -- in fact by the ``solitonic" NS5-brane of heterotic or type II theories wrapped on the $K_4$ part of the compactification $K_4\times T^2$.  Such an effective string arising from a wrapped NS5 brane is well known \cite{Polchinski:1998rr,Gauntlett:2003di}.   Integrating the NS5 $B_2$ flux over the boundary at $r\rightarrow \infty$ reproduces the integer winding number of the axiverse string. 

At scales $r\lesssim 1/m_\varphi$ set by the dilaton mass the dilaton is effectively no longer frozen
and the physics deviates from the inner-core solution as presented in Sec.\ref{sec:strings}.  Instead the
full 10d NS5-brane solution \cite{Polchinski:1998rr} becomes asymptotically operative as $r\lesssim {\rm min}(1/m_\varphi,R_4)$ where $R_4$ is the linear size of $K_4$.   (In other cases, involving ALPs from RR $k$-forms
the effective string at the core of the solution is a suitable wrapped D-brane.)

\section{\label{sec:general_strings}General Axiverse Strings}

We can generalise this analysis to large classes of axiverse strings.
The solution depends on the $G_{i{\bar j}}(m,{\bar m})$
metric behaviour in the limit $G_{i{\bar j}}\rightarrow 0$, as well as the potential $V(m,{\bar m})$.   We assume $V$ has three properties: a) There is a local minimum with $V\simeq 0$ at which all non-ALP saxion moduli, ${\rm Im}(m^i)$, are stabilised; b)  Stabilisation occurs at $\vev{{\rm Im}(m^i)} = b^i\gg1$ to have control over the 4d effective action; c) For ${\rm Im}(m^i)\rightarrow \infty,$ $V\rightarrow 0$.   

The string solution is then a semi-geodesic ``force-modified-motion" on the moduli space 
found by solving
\beq
{\boldsymbol\nabla}^2 m^i + \Gamma^i_{jk}(m) ({\boldsymbol\nabla} m^j).({\boldsymbol\nabla} m^k) = \partial_i  V(m)~,
\label{eq:semi-geodesic}
\eeq
subject to appropriate boundary conditions. Here $\Gamma$ is the connection derived from $G$.
The metric is most important in the string core where the effects of $V$ are
small.  

Turning to the moduli, in addition to the heterotic and type II ALPs arising from $B_2$ (Sec.\ref{sec:Axiverse}), type II strings on a CY orientifold, $Z$, also have ALPs arising from RR $k$-forms $C_k$.  As orientifolds now have fields and basis forms split in to even/odd sectors, for IIA the result is $h^{1,1}_{-}$ complex K\"ahler moduli $T^i$  defined by expanding $B_2 + iJ = T^i \omega_i = (b^i+ it^i) \omega_i$ where $\omega_i$ is a basis of $h^{1,1}_{-}$ odd (1,1)-forms and $J$ is the K\"ahler (1,1)-form of $Z$.  The saxion components $t^i$ measure the (dimensionless) size of 2-cycles of $Z$.   These are accompanied by $h^{2,1}+1$ complex structure (CS) moduli, including the axio-dilaton, with ALPs $\be^a$ defined by $C_3 = \be^a(x) \alpha_a$.  Here $\alpha_a$ are a basis of even harmonic 3-forms on $Z$ \cite{Svrcek:2006yi,Ibanez:2012zz}.

For IIB, there are $h^{1,1}_{+}$ K\"ahler moduli with ALPs $c^i$ defined by $C_4 = c^i(x)  {\tilde \omega}_i$ where ${\tilde \omega}_i$ are a basis of even $(2,2)$ forms dual to the even (1,1)-forms.  The associated saxion components $\tau^i(x)$ measure the size of 4-cycles of $Z$.  In addition there
are $h^{1,1}_{-}$ ALPs $d^i$ arising from $C_2 = d^i(x)\omega_i$, and also the 4d axio-dilaton field $S$ \cite{Cicoli:2012sz,Ibanez:2012zz}.

All the above are `closed-string' axions, but a wide variety of ALPs exist in Type I string theory and M-theory, and there often arise other potential axions too.   
A fraction of all these ALPs can be made massive at tree level by a variety of mechanisms, 
but some remain massless before non-perturbative effects finally lift them, and are candidates for associated axiverse cosmic strings.

Turning to the kinetic metric, consider, eg heterotic and IIA K\"ahler $T^i=b^i+ it^i$ moduli.  Since for a 6d CY
the volume can be written ${\cal V}(x) = \kappa_{ijk} t^i t^j t^k /6$ where $\kappa_{ijk} = \int_Z \om_i \wedge \om_j \wedge \om_k$ is the integer-valued triple intersection number, and the leading K\"ahler potential is of
form $K = -2\log({\cal V}+\dots)$ one finds for heterotic and IIA cases
\beq
G_{i{\bar j}}^{B_2}(x) = \frac{\kappa_{ikl} t^k t^l  \kappa_{jmn} t^m t^n}{16{\cal V}^2} -  \frac{\kappa_{ijk} t^k}{4{\cal V}}+\ldots 
\label{eq:B2metric}
\eeq
up to corrections which are small if $t^i \gg 1$.  
Note that $G^{B_2}(t^i \rightarrow \infty)\sim 1/t^2$ just like the $1/\rho_2^2$ scaling
of the $T^2$ metric. 
Up to a factor of the coupling $g_s^2$ the metric for the $C_2$ and 2-cycle $d^i$ ALPs in IIB is identical to eq.(\ref{eq:B2metric}), while for the $c^i$-moduli of IIB arising from $C_4$ and 4-cycles
\beq
G_{i{\bar j}}^{C_4}(x) = g_s^2 \left(\frac{ t^i t^j }{8{\cal V}^2} -  \frac{(\kappa_{ijk} t^k)^{-1}}{2{\cal V}}\right)+\ldots , 
\label{eq:C4metric}
\eeq
an implicit function of the 4-cycle (strictly, divisor)
volumes $\tau_i = \kappa_{ijk}t^j t^k/2$.  As $\tau^i \rightarrow \infty$, $G^{C_4}\sim 1/\tau^2$.

A simple two K\"ahler moduli example is the IIB orientifold of 
$P^4_{(1,1,1,6,9)}$ with
${\cal V}=(3t_1^2 t_5 + 18 t_1 t_5^2 + 36 t_5^3)/6$ where  $t_1,t_5$ are the 2-cycle volumes (in the notation of \cite{Denef:2004dm} where it was argued that all CS and dilaton moduli can be stabilised by fluxes, and later successful K\"ahler moduli stabilisation too \cite{Balasubramanian:2005zx}).   The 4-cycle volumes, the
partners of the $c^{4,5}$ ALPs from $C_4$, are
$\tau_4 = t_1^2/2$ and $\tau_5 =(t_1+6t_5)^2/2$ giving ${\cal V}=(\tau_5^{3/2}-\tau_4^{3/2})/9\sqrt{2}$. The $G^{C_4}$ metric is thus
\beq 
\frac{g_s^2}{36 {\cal V}^2} \begin{pmatrix}      
[2\tau_4+\tau_5^{3/2}/\tau_4^{1/2}]/3 & -(\tau_4\tau_5)^{1/2} \\
      -(\tau_4\tau_5)^{1/2} & [2\tau_5+\tau_4^{3/2}/\tau_5^{1/2}]/3 \\
   \end{pmatrix}.
\eeq
Now consider, eg, a string formed by the winding of $c^5$. 
The inner core behaviour is found by expanding the metric in the limit, $\tau_5\rightarrow\infty$, $\tau_4 = {\rm fixed}$, giving a diagonal metric with $G_{55}\simeq 3/\tau_5^2$. 
As this is the same $1/\rho_2^2$ scaling as for $T^2$,  and by definition the potential is unimportant in the inner core, the inner core behaviour of $m^5 \equiv c^5+i \tau^5$ derived from eq.(\ref{eq:semi-geodesic})
is identical to eq.(\ref{eq:innercoresol}) except the decompactification is to 8d!

In the far-field region the metric is effectively frozen at its asymptotic, stabilised value, so one again finds a $1/r^2$ 
dependence in the appropriate combination of moduli (upon diagonalising the asymptotic metric) as in
eq.(\ref{eq:farfieldsol}), and a tension of the same form as eq.(\ref{eq:T1tension}).  Only for the
detailed matching in the outer-core region does the full kinetic metric (and potential) need to be kept. 

As emphasised in Sec.\ref{sec:GravityWarping} in the deep inner core the axiverse string
fields are sourced by an effective `magnetic' string in the decompactified 8d theory.  In the
case of an ALP from the RR $C_4$ this string in 8d arises from a $D3$ brane wrapped over
the dual 2-cycle to $\tau_5$.


In all cases we have examined we have found qualitatively similar behaviour to that of Secs.\ref{sec:strings},\ref{sec:GravityWarping}, though we do not know if this covers all axiverse string possibilities.

\section{\label{sec:pheno}Phenomenology}

We now briefly discuss aspects of
axiverse string 
phenomenology \cite{future}.  The non-perturbative potential for the ALP itself has so far been ignored.    This gives it a tiny mass and it is energetically favourable for the ALP winding to fall into a domain-wall configuration. 
As strings now bound domain walls the overall network is unstable to decay.  Because of this such string-derived axion cosmic strings were excluded from consideration in \cite{Copeland:2003bj}.
Nevertheless the presence of the decaying string network can lead to important physical effects in the expanding universe.  In particular, as some of the axiverse ALP masses can be as small as $10^{-20}\eV$ or even $H_0 \sim 10^{-33}\eV$ (giving a favoured implementation of quintessence) \cite{Arvanitaki:2009fg}, the network can survive to the post-recombination epoch, or even the present. 
If the axion couples to electromagnetism via $aF{\tilde F}$ then strings can give rise to quantised polarisation rotation of CMBR photons, as well as zero modes mandated by index theorems and associated forms of superconductivity \cite{Witten:1984eb,Callan:1984sa,Manohar:1988gv,Alford:1990mk,Alford:1990ur,Brandenberger:1996zp,Agrawal:2019lkr,Agrawal:2020euj,Fukuda:2020kym}.  

As striking, the
inevitable \emph{long-distance} $1/r^2$ variation, eq.(\ref{eq:farfieldsol}), of the associated
non-ALP modulus ${\rm Im} (m^i)$
leads to new effects characteristic of axiverse strings.  As in string theory all SM couplings are set by  VEVs of moduli, variations of SM couplings such as Yukawa and gauge couplings in the vicinity of the string are to be expected.  
One variation that is essentially guaranteed is the mass of the ALP itself as almost always
the non-perturbative effects that set the potential $|{\tilde V}| \sim \exp(-S[ {\rm Im} (m^i)])$ depend exponentially on
${\rm Im} (m^i)$;  here $S$ is a suitable instanton action. Moreover SM Yukawa couplings of the light quarks and leptons in string theory also often depend \emph{exponentially} on moduli as they too are generated non-perturbatively \cite{Cremades:2003qj,Ibanez:2006da,Blumenhagen:2007zk}, so masses and mixings of the light SM fermions can vary strongly in the vicinity of the string.  
These changes can be either equivalence principle violating or preserving (in which case $G_N$ is 
effectively being rescaled) as will be elucidated in \cite{future}.

{\bf Acknowledgments:} We are grateful to Ed Hardy for comments. HT thanks the STFC for a
postgraduate studentship. 

\bibliography{axi_strings.bib}

\end{document}